\documentclass[final,3p,times,twocolumn]{elsarticle}

\usepackage{amssymb}
\usepackage{soul}
\usepackage{amsmath}

\usepackage{float}
\usepackage{xcolor}
\definecolor{RED}{RGB}{156,78,90}

\journal{Journal of Colloid and Interface Science}

\begin{document}

\begin{frontmatter}



\title{On evaporation dynamics of an acoustically levitated multicomponent droplet: evaporation-triggered phase transition and freezing}


\author[1]{Hao Zeng}
\address[1]{Center for Combustion Energy, Key Laboratory for Thermal Science and Power Engineering of Ministry of Education, Department of Energy and Power Engineering, Tsinghua University, Beijing 100084, China}

\author[1]{Yuki Wakata}

\author[1]{Xing Chao\corref{cor1}}
\ead{chaox6@mail.tsinghua.edu.cn}

\author[1]{Mingbo Li\corref{cor1}}

\ead{mingboli@mail.tsinghua.edu.cn}

\author[1,2]{Chao Sun\corref{cor1}}
\ead{chaosun@tsinghua.edu.cn}

\address[2]{Department of Engineering Mechanics, School of Aerospace Engineering, Tsinghua University, Beijing 100084, China}
\cortext[cor1]{Corresponding author}

\begin{abstract}
\textbf{Hypothesis:}\\
Multi-component droplet evaporation has received significant attention in recent years due to the broad range of applications such as material science, environmental monitoring, and pharmaceuticals. The selective evaporation induced by the different physicochemical properties of components is expected to influence the concentration distributions and the separation of mixtures, thereby leading to rich interfacial phenomena and phase interactions.\\
\textbf{Experiments:}\\
A ternary mixture system containing hexadecane, ethanol, and diethyl ether is investigated in this study. The diethyl ether exhibits both surfactant-like and co-solvent properties. Systematic experiments were performed using acoustic levitation technique to achieve a contact-less evaporation condition. The evaporation dynamics and temperature information are acquired in the experiments, using high-speed photography and infrared thermography technologies.\\ 
\textbf{Findings:}\\
Three distinct stages, namely, `Ouzo state', `Janus state', and `Encapsulating state', are identified for the evaporating ternary droplet in acoustic levitation. A self-sustaining periodic freezing \& melting evaporation mode is reported. A theoretical model is developed to characterize the multi-stage evaporating behaviors. We demonstrate the capability to tune the evaporating behaviors by varying the initial droplet composition. This work provides a deeper understanding of the interfacial dynamics and phase transitions involved in multi-component droplets and proposes novel strategies for the design and control of droplet-based systems.
\end{abstract}




\end{frontmatter}


\section{Introduction}

Evaporation of liquid droplets is a ubiquitous phenomenon in daily life and has implications in various areas~\cite{lohse2020physicochemical}, such as ink-jet printing~\cite{park2006control,yoo2015experimental,lohse2022fundamental}, spray cooling~\cite{jia2003experimental,liang2017review}, particle assembly~\cite{tchakalova2014evaporation,tan2019porous}, and biomedical productions~\cite{wong2014evaporative,bui2017ordered}. Because of such a wide range of industrial applications, this phenomenon has been under investigation for many years and has attracted increasing research interest in the field of soft condensed matter and fluid mechanics. 

The fundamental physics of evaporation is based on the transport of mass and heat from the surface of a liquid to the surrounding gas phase. Single-component liquid droplets offer an ideal platform to understand the core physical mechanism of this transport phenomenon as they have well-defined surfaces and homogeneous physical properties. The process of evaporation is controlled by the diffusion of vapor molecules from the surface of a droplet to the surrounding air, in which the ambient conditions such as temperature~\cite{kincaid1989water}, pressure~\cite{kitano2014effects}, and surrounding vapor concentration~\cite{pinheiro2019ethanol} play a crucial role. In the liquid phase, key factors include surface tension gradient~\cite{niazmand1994effects}, viscosity~\cite{lu2011internal}, and density gradient~\cite{Edwards2018density}, which enhance or depress the major causes. The thermodynamics of evaporation meet mechanics on the droplet surface to result in rich flows and mass transfer inside the evaporating droplet.

In reality, most liquid droplets are not composed of a single component. The presence of different chemical species in a droplet can lead to complex physical and chemical behavior\cite{ravindran1982multicomponent, RUBEL1981188}. Extensive studies on the problem of multi-component liquid droplet evaporation have shown that the variable liquid properties and interactions between the liquid components can not be neglected in predicting the accurate evaporation process. Abraham and Magi~\cite{abramzon1989droplet} modeled the multi-component droplet vaporization in sprays, pointing out the importance of the volatility of different components for evaporation. Brenn et al.~\cite{brenn2007evaporation} presented a realistic model for describing multi-component liquid droplet evaporation of up to five components. Their model shows the necessity to consider the contributions of each individual component to the evaporation and the interactions between the liquid components. The difference in volatility between components would lead to the selectivity of the evaporation process, in which the most volatile components evaporate the fastest~\cite{jeong2021selective}. As a result, the selective evaporation mode causes peculiar differences in the composition of the mixture~\cite{de2021marangoni}.

More interestingly, for some multi-component liquid systems, the evaporation-triggered concentration change may give rise to the spontaneous phase transition behaviors and complex flows~\cite{zang2019evaporation, MANDAL2012260}. For example, the well-known "Ouzo effect", in which tiny droplets of poor molecular solubility component spontaneously nucleate in bulk and lead to liquid-liquid phase separation of the multi-component mixture~\cite{vitale2003liquid}. The existence of phase transitions and the coupling with mechanical instabilities~\cite{sternling1959interfacial,grahn2006two,de2021marangoni} could result in rich droplet dynamics including self-seal~\cite{tan2016evaporation}, explosion~\cite{lyu2021explosive} and self-arrange into droplet patterns~\cite{lopez2021phase}. In some "smart" systems, the mixture can self-organize during the course of liquid evaporation to form and finish stable compartmentalization~\cite{yuan2017phase,guo2021non} and macro-segregation~\cite{chen2022evaporation} by phase separation. 

Fundamentally, the underlying physics of multi-component droplet evaporation has still not been comprehensively understood. This situation is particularly exacerbated by the presence of a tri-phase contact line (TCL)\cite{VANGAALEN2020888}, where the three phases of liquid, solid, and vapor meet. The evaporation profile of sessile droplets exhibits non-uniformity at the triple line~\cite{hu2020evaporation}. This non-uniformity further influences the mass transport inside the evaporating droplet, leading to complex inner flow structure~\cite{diddens2017modeling, VANGAALEN2021622} and preferential solute deposition on the substrate~\cite{deegan1997capillary}. To suppress the effects of TCL on evaporation, various levitation techniques are applied to investigate droplet evaporation. Among them, acoustic levitation is one of the most powerful techniques for studying drop dynamics due to its strong levitation capacity and wide applicability to various samples. The acoustic field can create steady annular streaming around the evaporating droplet~\cite{trinh1994experimental}. A theoretical description is provided by Yarin et al.~\cite{yarin2002evaporation} to interpret the convective effect of the acoustic streaming arising near the free droplet surface. Chen et al.~\cite{chen2022evaporation} reports the occurrence of liquid phase separation and water condensation of acoustically levitated evaporating binary droplets. The nonlinear effect and convective streaming in the acoustic field can provide unique boundaries to evaporation and have attracted increasing research interest.

In this study, we experimentally investigated the evaporation dynamics of a three-component droplet. The ternary droplet is levitated in an acoustic field to avoid contact with any other substrates~\cite{zang2019evaporation, sasaki2020heat,andrade2018review}. We essentially focus our attention on exploring the rich phenomenology triggered by differences in the physicochemical properties of individual components of the droplet. Therefore, we chose fluid components with significant differences in physicochemical properties to construct the ternary droplet system. With a composition of diethyl ether, ethanol, and hexadecane, unexpectedly, the life of the droplet exhibits three distinctive stages, involving rich evaporation-triggered inter-phase behaviors. Interestingly, spontaneous solidification of the liquid droplet is observed in an atmospheric environment without contact with any cold source. Following the pioneer's works~\cite{brenn2007evaporation,al2011evaporation}, a multi-stage model is developed by considering the influence of solidification layer and droplet configurations to predict the droplet size and composition evolution in the evaporation process. Finally, we present distinct features in the entire life process of the ternary droplets according to different initial compositions. The findings may have instructive meaning to applications utilizing the evaporation and solidification of complex liquids.

\section{Experimental Section}

\subsection{Materials}
                       
Ethanol (General-Reagent, $\ge 99.7\%$), diethyl ether (Sinopharm Chemical Reagent Co. Ltd, $\ge 99.7\%$), and hexadecane (Macklin, 99\%) are used in the experiment to prepare the ternary mixture. For the current formula, the diethyl ether is completely and mutually soluble to the other two components while ethanol is only slightly soluble in hexadecane. The saturated vapor pressure of diethyl ether and ethanol is 71.5 kPa and 7.9 kPa at room conditions (25$^{\circ}$C, 1 atm), respectively, emphasizing that these two components are quite volatile, especially for diethyl ether. Hexadecane exists as a liquid and can hardly evaporate at room conditions but with a melting point of 18$^{\circ}$C. The system composition is comparable to a general ternary liquid system. The detailed information on the three components at 25$^{\circ}$C is listed in Table.~\ref{composition1}. All chemicals were used without further purification. The ternary solution is prepared by dropping diethyl ether into the pre-blended binary mixture of hexadecane/ethanol in cryogenic vials at 25$^{\circ}$C. The vials are then placed on a vortex oscillator (Kylin Bell, Vortex-5) to fully mix.

\begin{table}
\small
  \caption{Physical properties for three pure components at 25$^{\circ}$C. ( $\rho$: density, SVP: saturated vapor pressure, $T_m$: melting point temperature.)}
  \label{composition1}
  \begin{tabular*}{0.48\textwidth}{@{\extracolsep{\fill}}lllll}
     \hline
    \textbf{Component}&\textbf{Formula}&$\rho$($\rm kg/m^3 $)&SVP(kPa)&$T_m$($^{\circ}$C)\\
    \hline
Hexadecane& $\rm C_{16}H_{34}$ & 769.9 & 0.0002 & 18.0\\
Ethanol& $\rm C_{2}H_{5}OH$ & 785 & 7.9 & -114.1\\
Ether& $\rm C_{2}H_{5}OC_{2}H_{5}$ & 708 & 71.5 & -116 \\
    \hline
  \end{tabular*}
\end{table}

\subsection{Experiment setup}

\begin{figure}
\centering
\includegraphics[scale = 0.9]{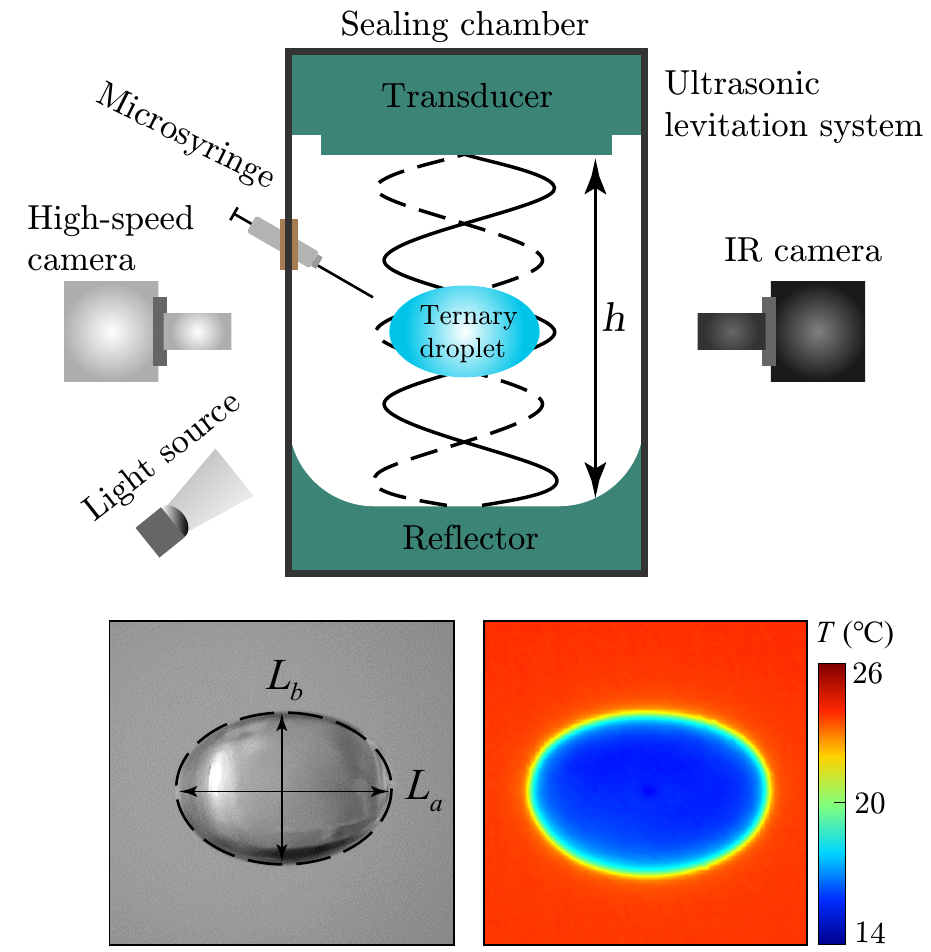}
\caption{Schematic of the experimental setup. The two images at the bottom display the typical images of the ternary droplet at the evaporation initiation obtained by a high-speed camera and infrared camera respectively.}
\label{FIG1}
\end{figure}

The experiment setup is shown in Fig.~\ref{FIG1}. An optically transparent chamber is mounted surrounding the levitator. We conducted the experiments in a closed chamber to avoid possible perturbations from ambient airflow. A coaxial acoustic levitator is utilized in the experiment to levitate the droplet. The acoustic levitator consists of two parts: the transducer and the reflector. The transducer is attached to a piezoelectric crystal to generate ultrasonic waves at a fixed frequency, i.e., in this study, 20.5 kHz. The reflected sound wave interferes with the emitted wave to form a standing-wave acoustic field (four pressure nodes in this experiment). After the formation of standing waves, the sound pressure in both the upward direction and the downward direction is generated and distributed sinusoidally. The net force exerted by the sound pressure is zero at the nodes, so the equilibrium position for the droplet is slightly below a pressure node against gravity, where the net pressure force is upward. The balance between the acoustic field and gravity leads to non-uniform pressure distribution along the droplet surface and thus once released, the droplets appear ellipsoidal. By adjusting the distance $h$ between the transducer and reflector, we are able to manipulate the droplet shape and ensure the best levitating stability. 

Before any experiment, the ternary liquid has been sealed in the injection channel and isolated from the air. The environment temperature is controlled by air conditioning facilities to be constant at 25$^{\circ}$C. The levitation apparatus provides a substrate-free condition for the experiments to investigate the droplet evaporation and solidification behaviors. In the experiment, a ternary droplet of a certain volume is manually injected near one of the acoustic pressure nodes using a microsyringe. The initial volume $V_0$ of the droplet is $2.4\pm 0.15\ \mu \rm{L}$ with a corresponding equivalent diameter $D_{eq}=(6V_0/\pi)^{\frac{1}{3}}$ of about 1.7 mm, which is below the capillary length scale. The inject position is carefully determined to achieve equilibrium as quickly as possible and prevent droplet drying before recording.

\begin{figure*}
\includegraphics[width = 0.98\textwidth]{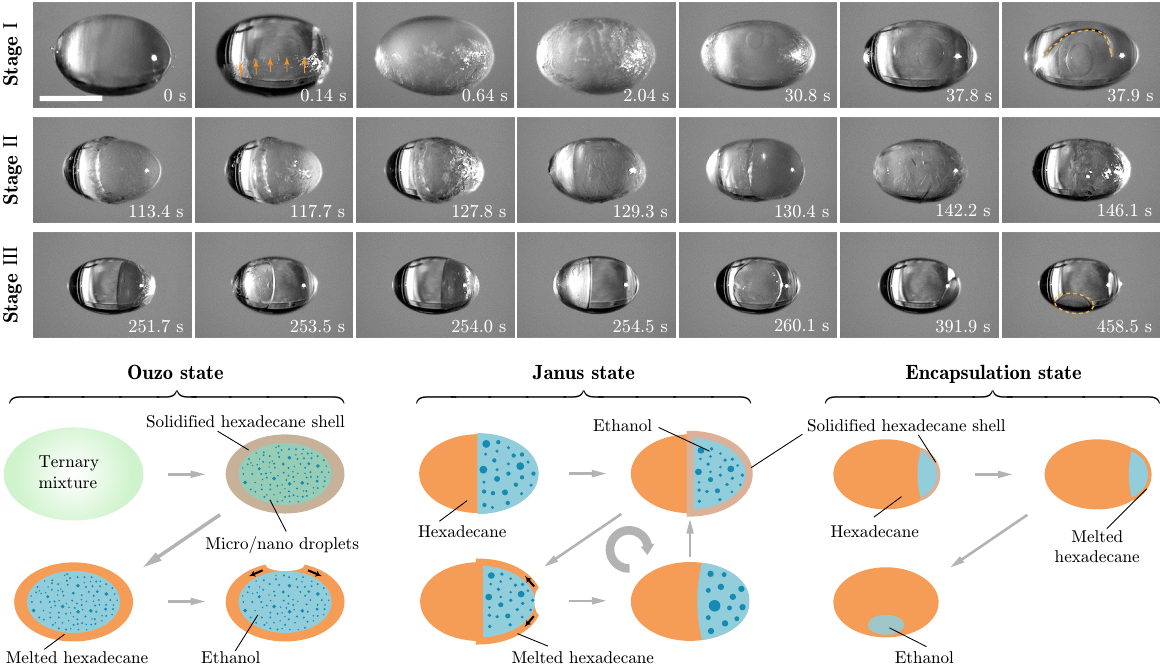}
\caption{Entire lifetime of an acoustically levitated ternary droplet. \textbf{Top:} the side-view snapshots of the evaporating droplet captured by a high-speed camera. The origin and direction of the arrows mark the instantaneous location and direction of the freezing front. \textbf{Bottom:} representative schematic detailing three typical stages that the ternary droplet goes through during evaporation (not to scale). The moment '0 s' means that the droplet has just been released from the tip of the needle into the air. Snapshots are not selected at equal intervals in order to comprehensively demonstrate the characteristic behaviors of the droplet in different stages. The recording was performed at 25$^{\circ}$C. The scale bar is 1 mm.}
\label{FIG2}
\end{figure*}

An imaging system, consisting of a high-speed camera (Photron Nova S12) and a long-distance microscope (Navitar Zoom 6000), is used to record the entire evaporation process of the ternary droplet. The high-speed camera is placed on the side and equipped with backlight illumination. The experimental images are recorded at a frame rate of 50 fps with a shuttle speed of 0.25 ms. The temperature distribution within the droplet is acquired by an infrared thermography system consisting of an IR camera (Telops FAST L200, Stirling-cooled MCT detector) and a 50 mm germanium objective with a micro extender ring. Here we use the instrument's built-in radiometric temperature measurement to convert the infrared images into temperature information. The calibrated temperature measurement range is from -40$^{\circ}$C to 150$^{\circ}$C, in which the measurement error can be guaranteed to be less than 1 $^{\circ}$C. The video and thermal imaging information are recorded once the droplet is released in the acoustic field.

\section{Results and discussion}
 
\subsection{Evaporation-triggered solidification of an acoustically levitated ternary droplet}


For the ternary mixture, the diethyl ether exhibits both surfactant-like and co-solvent properties, namely, dual actions in bulk solutions and interfaces. It dramatically increases the solubility of poorly soluble hexadecane in ethanol. With the presence of a high freezing point component, the phase transition from liquid to solid is introduced to the system. In a sealed environment with constant temperature and pressure, the ternary mixture is thermodynamically stable. Once it is exposed to the air, the huge difference in physicochemical properties between the components can induce an unexpected and complex phenomenon. 


As soon as the ternary drop is placed near the pressure node of the acoustic field, it automatically moves to the equilibrium position where the sound pressure force can balance gravity. The volume of the droplet contracts as soon as the droplet exposes to the surrounding air and starts to evaporate. Figure~\ref{FIG2} shows representative images of temporal slices taken from a side-view high-speed recording of the evaporating droplet. Note that due to the non-linear effect presented in the evaporation dynamics, the sequence presented here is not arranged at equal time intervals. The entire evaporation process can last for dozens of minutes, with three distinct stages experienced. 

Initially, the droplet is macroscopically homogeneous, stable, and transparent, which means it is in the one-phase regime and all components are molecularly miscible. Due to the large saturation pressure at room temperature, diethyl ether preferentially evaporates rapidly upon exposure. The effect of diethyl ether evaporation on the droplet body is twofold. On the one hand, a large amount of latent heat is released that causes the droplet surface to become supercooled within milliseconds. On the other hand, the change in the composition of the ternary droplet inevitably triggers phase separation and component diffusion, especially near the surface. A large number of discrete micro/nano-droplets precipitated instantaneously (0.14 s). The droplet appearance gradually changes from transparent to cloudy, a phenomenon that is called the "Ouzo effect"~\cite{sitnikova2005spontaneously}. The "Ouzo effect" is one of the most distinctive phenomena during the evaporation of multi-component droplets and is observed in many similar studies~\cite{tan2016evaporation,lyu2021explosive, li2022spontaneously, lopez2021phase}. Thereafter we refer to the evaporating ternary droplet in the first stage as the "Ouzo state". In this state, the metastable Ouzo droplets ($\sim\mu$m) always coexist with nanoscale aggregates (a recently confirmed phenomenon with the terminology 'pre-Ouzo', surfactant-free microemulsions or mesoscale solubilization) present in the bulk in dynamic equilibrium~\cite{zemb2016explain, schoettl2014emergence, prevost2021spontaneous}.    

As the dispersed micro/nano-droplets are full inside the droplet, it was found that solidification occurs from the nucleation sites on the surface of the droplet (0.64 s). The melting point of the non-volatile component hexadecane (18$^\circ$C) is just slightly below room temperature (25$^\circ$C), which means it is easy to solidify under the evaporative cooling effect. The solidification event initiates from the southern hemisphere of the droplet, and propagates rapidly to cover the entire surface. Here the temperature distribution on the droplet surface caused by evaporative cooling may not be uniform, and in the experiments the initial site of solidification was random. Moreover, the recalescence phenomenon that occurs in the freezing of supercooled water droplets has not been observed in this experiment. It suggests that there is a completely different physical picture within the evaporation-triggered solidification.

After the droplet surface is fully covered with a layer of solidified hexadecane, the position of the droplet moves slightly upward due to the better reflection of the acoustic wave of the solid surface, and the droplet deforms from an ellipsoidal shape into a thin spherical disk as a result of the elevated acoustic pressure~\cite{zang2018inducing}. The changes in droplet shape show a lack of rigidity to suppress the acoustic deformation of the levitated droplet. In contrast to the freezing of supercooled one-component droplets, solidification only occurs on the droplet surface and an interconnected stable frame is absent under this case~\cite{bauerecker2008monitoring}. The inner flow of the droplet is suppressed with the solid shell covering the droplet surface. The solid layer insulates the droplet from the convective outer acoustic streaming and imposes a no-slip boundary condition on the droplet surface. The surface freezing quickly encloses the droplet forming a solid hexadecane shell (2.04 s). Without effective evaporative cooling, the solid hexadecane shell eventually melts a few decades of seconds later (37.8 s). A thin liquid film wrapping the inner mixture can be observed. Once directly exposed to the outer acoustic streaming again, this covering hexadecane film rapidly ruptures due to the strong perturbation introduced by the convective boundary. Here we define the moment of film rupture as the end of the Ouzo state. It is difficult to assess whether diethyl ether has completely been depleted at this time due to the complex interaction between volatile components.

Interestingly, the melted hexadecane gradually gathers, enabling the formation of a multicompartmental droplet, one hexadecane-rich, and one ethanol-rich. As shown in the middle row of Fig.~\ref{FIG2}, the hexadecane-rich phase remains transparent, while the ethanol-rich phase is turbid, in which a large number of hexadecane micro/nano-droplets are suspended. The two exposed segments are horizontally bonded together with a clear interface (113.4 s), very similar to the configuration of the so-called 'Janus droplet'~\cite{yuan2017phase, guo2021non}. The Janus droplets are compound droplets that consist of two adhering drops of different fluids that are suspended in a third fluid. Here we refer to the second stage in the lifetime of the evaporating ternary droplet as the 'Janus state'. Surface tension is the dominant driving force for the conversion to Janus droplet. For the ethanol-enriched compartment, the semi-ellipsoidal droplet cools down again due to the evaporation of remaining volatile components. A solid layer is preferentially formed at the interface between the two phases to further stabilize this Janus droplet configuration. Then a semi-ellipsoidal solid hexadecane shell formed, enveloping the ethanol-enriched mixture. Importantly, we noticed that in this stage, solidification and melting occurred alternately on the surface of the ethanol-based compartment, as shown in the experimental snapshots (113.4s$\sim$146.1s) in Fig.~\ref{FIG2}. This periodic "solidification-melting" behavior is observed in the entire stage when the droplet is configured as a Janus droplet. The volume of the ethanol-enriched compartment noticeably decreases due to the evaporation of ethanol. Meanwhile, part of the hexadecane dissolved in this compartment is accumulated from the bulk to the surface by solidification and then aggregates into the hexadecane-enriched compartment after melting.






\begin{figure*}
\includegraphics[width = 0.98\textwidth]{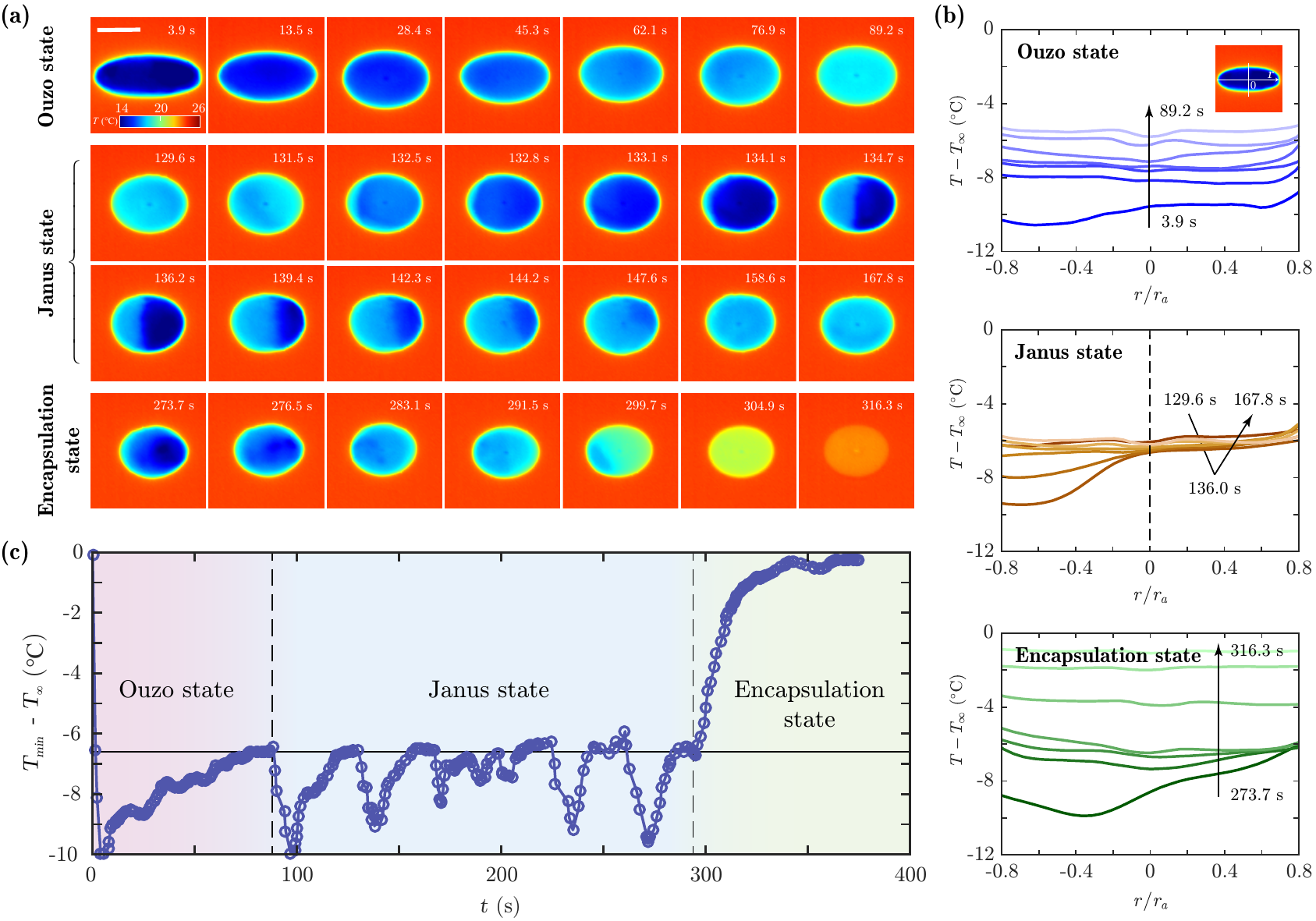}
\caption{Temperature distribution of the ternary droplet during evaporation under room-temperature 25$^\circ \rm C$ conditions. (a) Time-lapse thermographic images. Time '0' corresponds to the ternary droplet just being exposed to the air. (b) Temporal evolution of temperature profiles along the equator of the droplet in the three stages. The profiles are temporally equally spaced in each figure and with colors fading with time. $T_\infty$ represents the environment temperature, and $r_a$ is the long axis radius of the droplet. (c) Temporal evolution of the lowest temperature within the ternary droplet during the evaporating. The solid line represents the melting temperature of hexadecane.}
\label{FIG3}
\end{figure*}

Upon further decrease of the volume of the ethanol-based compartment, the solid hexadecane shell becomes quite thin and can even not cover the entire surface of the compartment. Finally, the "solidification-melting" cycle failed and the third stage is launched. In this stage, the melted hexadecane predominates to wrap the ethanol-based phase, thus inhibiting evaporation. We refer to the ternary droplet in this stage as the so-called "encapsulation state"~\cite{IQBAL2015173,tu2014one}. The snapshots in the third row of Fig.~\ref{FIG2} show that the encapsulated biphasic configuration exists stably with an easily identifiable interface. The volatile component inside, namely ethanol, is slowly dissolved into the outer phase to compensate for the loss caused by evaporation through the gas-liquid interface. As ethanol is slightly soluble in hexadecane, the evaporation rate is dramatically reduced in this "encapsulation state". Note that this stage lasts for several ten minutes and occupies most part of the time in the entire evaporation process. This extremely slow activity is not fully shown. After this point, the droplet turns into a monophasic state with only hexadecane remaining.

\subsection{Temperature field of the evaporating ternary droplet}

As known to all, the surface evaporation profile is strongly coupled with the local temperature distribution. By utilizing a high-precision infrared camera (Fast L200 by Telops Inc.), we are able to track the evolution of temperature distribution on the surface of the droplet over its entire lifetime, as shown in Fig.~\ref{FIG3}. Within the response wavelength band of the detector from 7.7 to 9.3 $\mu$m, the solidification process on the droplet surface has a negligible effect on the thermal emissivity in analogous to ice/water system~\cite{hori2006situ}. In fig.~\ref{FIG3}(b), the temperature profile evolutions on the droplet equator in different stages are presented, in which the horizontal positions are normalized by the long axis radius $r_a$ of the droplet. The absolute temperature value is subtracted by the average of the background (environment) to compensate for measurement error. And the temperature information near the edge of the droplet is not presented to minimize the effect of the angular dependence of thermal emissivity on the result~\cite{rees1992angular}.

It can be observed that the temperature field within the droplet shows distinct characteristics at the three stages. As shown in the previous section, once the droplet is released, micro hexadecane drops precipitate out from the continuous phase and are cooled by the rapid evaporation, initiating freezing at a temperature below 18$^\circ$C on the liquid-gas interface. We start from the moment when the droplet has been wrapped in a layer of solid hexadecane shell ($t = 3.9$ s). After the droplet surface is completely frozen, the evaporative-cooling effect fails and the temperature in the solid phase turns to rise progressively until the end of the Ouzo state (see Fig.~\ref{FIG3}(b)). There appears to be a nearly uniform temperature field on the surface of the solid shell during melting, as revealed by the uniform single color. Note that here we only obtain the temperature distribution on the droplet surface, not the region inside. There should be a noticeable temperature gradient inside the droplet due to the intense diffusion and thermal convection~\cite{kneer1993diffusion,lyu2021explosive}. As shown in Fig.~\ref{FIG3}(c), the global minimum temperature, With regard to this relatively homogeneous temperature distribution, the global minimum temperature can well characterize the solidification and melting process on the evaporating droplet surface.

In the Janus state, evaporation-triggered freezing and melting occur in sequence as the temperature falls and then rises. We show a complete freezing $\&$ melting cycle of the Janus state in Fig.~\ref{FIG3}(a) and (b). The ethanol-rich compartment undergoes evaporation more violently, which leads to a higher temperature reduction and more probable nucleation. A sharp temperature gradient can be observed near the phase boundary inside the droplet(134.7 s). Whereafter, after reaching the minimum temperature of about 14.5$^\circ$C, the solid hexadecane shell starts to melt naturally with a gradual increase of the temperature. The hexadecane-rich compartment, though it also cools, has a surface temperature fluctuation of less than 1$^\circ$C (see Fig.~\ref{FIG3}(b)). This indicates that the evaporation of residual volatile components dissolved in hexadecane is obviously more moderate compared with the other side. The end of this freezing $\&$ melting cycle is marked by the disappearance of the temperature gradient within the droplet, and the uniform temperature field on the surface is restored again. These two parts of the droplet undergo similar courses during the freezing $\&$ melting cycles. We find that seven freezing $\&$ melting cycles, with an average duration of $\sim$29 s, can be clearly identified in the Janus state (see Fig.~\ref{FIG3}(c)). The seemingly irregular fluctuations in the middle of the Janus state are due to the unstable rotation of the levitated droplet in the acoustic field, resulting in the ethanol-rich compartment facing the back and out of the IR camera's view.

With the end of the last freezing $\&$ melting cycle ($\sim$295 s), the droplet enters the encapsulation state, where the temperature distribution tends to be uniform and gradually approaches room temperature at a slow rate. Thereafter the temperature within the droplet almost remains constant. The slight temperature difference from the environment temperature is attributed to the limited effect of evaporative cooling of the slightly dissolved ethanol in the bulk hexadecane.

\begin{figure}[th]
    \includegraphics[width=0.45\textwidth]{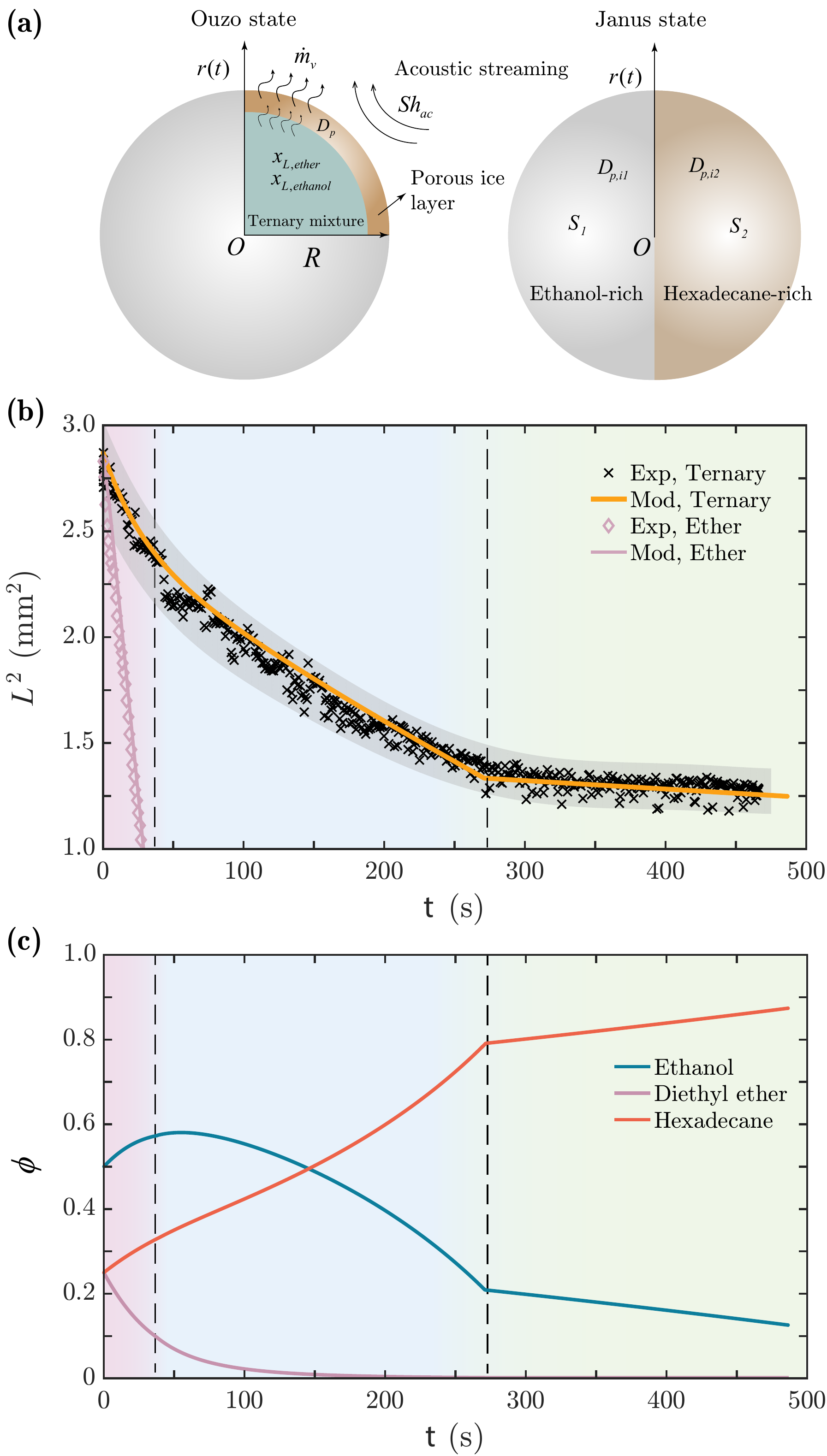}
\caption{Evaporation dynamics of the ternary droplet. (a) The ideal model of an evaporating droplet in the first and second stages (not to scale). (b) Comparison between the modeled effective evaporation area $L^2$ of the ternary droplet and experimental results. The experimental data (black cross) corresponds to the visual observation shown in Fig.~\ref{FIG2}. The scattering of the data originates from the droplet deformation, which inevitably leads to errors in identifying the droplet volume from the side-view snapshots. (c) Evolution of the droplet volumetric composition versus time. The volume fraction $\phi$ of each independent component is obtained from the model prediction.  }
\label{FIG4}
\end{figure}

\subsection{Long-time dynamics of the evaporating ternary droplet}

In this section, we will focus on the ternary droplet's evaporation dynamics, most importantly, its evaporation rate. In the current set of experiments, the droplet can be approximated as a spheroid with its rotational symmetric axis in the vertical direction under the coupling effect of gravity and acoustic streaming. We introduced the volume equivalent diameter, $L=\sqrt[3]{L_b L_a^2}$, where $L_b$ and $L_a$ are the minor and major axes on the side-view of the droplet~\cite{yarin1999evaporation}, as the characteristic length scale. With progressive evaporation, the dynamics are characterized by the transient variation in effective evaporation area $L^2$, as shown in Fig.~\ref{FIG4}(b) (Black cross). The data scattering originates from the droplet deformation and fluctuation in the acoustic field. As expected, the variation of droplet volume is nonlinear in nature. Combined with the phenomenological description mentioned above, three distinct stages can be clearly identified within the entire evaporation process. With a rapid variation in the concentration of volatile species, the classical $D^2$-law becomes increasingly inaccurate in predicting the evaporation rate of the droplet. Correspondingly, we developed a multi-stage theoretical model to document the temporal evolution of droplet surface $L^2$ (orange curve), in which the influence of complex droplet configuration and interaction with the solid phase is captured. 
The conceptual sketch of the physical model is shown in Fig.~\ref{FIG4}(a). By assuming a spherical symmetry in the evaporation of pure liquid droplets~\cite{abramzon1989droplet}, the mass transfer process is described based on the film theory developed by Bird et al.~\cite{Bird1960}, in which the resistance to mass transfer is attributable to a stagnant, thin region at the interface between the liquid and gas phases. Applying Fick's law to the thin film surrounding the droplet, the mass flow rate of a pure droplet can be expressed as: 
\begin{equation}
    \label{eqn:mass}
    \dot m_v =2\pi R\rho_g D_g Sh \left(\frac{Y_{v,s}-Y_{v,\infty}}{1-Y_{v,\infty}}  \right),
\end{equation}
where $R$ is the radius of the droplet, $\rho_g$ is the density of the gas phase, $D_g$ is the diffusion coefficient of the vapor in the gas film, $Sh=2R(\partial Y/\partial r)|_s (Y_\infty-Y_s)$ is the Sherwood Number, and $Y_{v,s}$ and $Y_{v,\infty}$ are the mass fractions of the vapor at the droplet surface and in the undisturbed environment medium, respectively.
For a multi-component droplet, the evaporation can be calculated as the summation of the evaporation of each individual components~\cite{brenn2007evaporation}, as:
\begin{equation}
    \dot m =\sum_{i=1}^{N}\dot m_i=2\pi \sum_{i=1}^{N}R_i\Bar{\rho} \bar{D_i} Sh_i B_{M,i}
    \label{eq:diff}
\end{equation}
where $N$ is the total number of component species, $\dot m_i, R_i,$ and $\overline{D_i}$ is the mass flow rate, partial equivalent radius, and mean mass diffusion coefficient in the dry air for the $i$-th species, respectively. The partial radius is found to follow the 1/3-scale of the volume fraction in multi-component liquid to represent the evaporation rates of the mixture component and is defined as $R_i=R\phi_i^{1/3}$~\cite{brenn2007evaporation}, where $\phi_i$ is the volume fraction of the corresponding component. The Sherwood number $Sh_i$ and Spalding mass transfer coefficient $B_{M,i}$ are calculated with the vapor properties of each individual component. The component activities in the liquid mixture are also considered to ensure a realistic prediction of the gas phase mass fractions. The mass fraction on the droplet surface $Y_{v,s,i}$ is computed with the mole fraction $x_i$ which is calculated as:
\begin{equation}
    x_i=\frac{p_{sat,i}}{p_m}\gamma_ix_{L,i}
\end{equation}
where $x_{L,i}$ is the mole fraction in the liquid phase, $p_{sat,i}$ is the saturated vapor pressure of the $i$-th component, $p_m$ is the total pressure of the vapor mixture (which equals the atmospheric pressure), and $\gamma_i$ is the activity coefficient of the component in the liquid mixture, which is determined by the UNIFAC method~\cite{gmehling1982vapor}. For the calculations in our work, we use a version of UNIFAC software available on Matlab Central~\cite{Saeed2023unifac}. 
However, for an acoustically levitated droplet, the acoustic streaming significantly changes the evaporation boundary. Yarin et al.~\cite{yarin1999evaporation} develop a new correlation that defines an averaged Sherwood number in the acoustic boundary layer by considering the sound wave properties and droplet shape. It reads:
\begin{equation}
    Sh_{ac}=K_{ac}\frac{A_{0e}}{\rho_g c \sqrt{\omega_{ac} D_g}},
\end{equation}
where $K_{ac}$ is a general factor that relates to the droplet shape and can be approximate to 1.89 for small droplet~\cite{yarin1999evaporation}, $\omega_{ac}$ is the frequency of the sound wave, $c$ is the speed of sound, and $A_{0e}$ is the effective sound wave amplitude and can be defined by a correlation with the effective sound pressure level (SPL)~\cite{yarin1998acoustic}, which is approximately 94.5 in our experiment. 
As shown in Fig.~\ref{FIG4}(a), the solidified hexadecane layer is considered as a porous layer that enables the inner liquid to evaporate and permit through the micro-poles. The pore structures formed in the evaporation-solidification process are assumed to be uniform-sized and greater than the mean free path of the evaporated gas molecules. The Dusty Gas Model (DGM) is utilized to describe the diffusion process in porous media. Therefore, the mass diffusion coefficient $\bar{D}_{i}$ in Eq.~\eqref{eq:diff} can be modified by the expression~\cite{mason1967flow}: $\bar{D}_{p,i}=\epsilon \bar{D}_{i} / \tau$ where $\epsilon$ and $\tau$ are the porosity and the tortuosity factor for the porous medium, respectively. The values of $\epsilon$ and $\tau$ are empirical parameters that are related to the pore structure of the exact porous solid layer. Considering a temporal average effect of the solid layer on droplet evaporation, our experimental data shows good agreement with a constant ratio of $\epsilon / \tau=0.5$, and the final model predictions are consistent with all cases considered in this study.

In Stage 1, the droplet surface is covered by the solid hexadecane layer most of the time. Thus, the exposure of gasified liquid to the surrounding environment is limited by the solid hexadecane layer. The mass flow rate in this stage can be given by:
\begin{equation}
    \dot m_1 =\sum_{i=1}^{N}\dot m_{1,i}=4\pi \sum_{i=1}^{N}R_i\Bar{\rho} \bar{D}_{p,i} \rm{ln}\it{(1+B_{M,i})},
\end{equation}
where the variables have the same definitions as previously clarified in Eq.~\ref{eq:diff}.
In Stage 2, the component concentrations vary greatly in different compartments of the Janus droplet. The mass flow rate is calculated as the sum of two individual compartments, while the evaporation is assumed to be uniform on a single phase surface, thus the mass flow rate in this stage can be calculated as:
\begin{equation}
    \dot m_2 =\sum_{j=1}^{2}\sum_{i=1}^{N}\dot m_{2,ij}=4\pi \sum_{j=1}^{2}s_j\sum_{i=1}^{N}R_{ij}\Bar{\rho_j} \bar{D}_{p,ij} \rm{ln}(1+\it{B_{M,ij})},
\end{equation}
where the subscript $i$ denotes the component species. As illustrated in Fig.~\ref{FIG4}(a), $j=1,2$ denotes the two separated compartments, and $s_j$ is the surface area ratio of the $j$-th compartment that can be determined with the volume fraction occupied by the compartment $j$.
In Stage 3, the liquid surface is directly exposed to the surrounding acoustic streaming. Thus the influences of the acoustic field need to be considered in this stage. The contribution of residual diethyl ether to the evaporation rate can be neglected in this stage and ethanol is the only volatile component. The mass flow rate in this stage is given by:
\begin{equation}
    \dot m_{3}=2\pi R_i\Bar{\rho_3}\bar{D} Sh_3 B_{M,3}.
\end{equation}
With given initial conditions and phase transition moments, this model can accurately capture the evolution dynamics at different stages. For a detailed derivation of the model, please refer to the Supplementary Materials.

\begin{figure*}[h]
\includegraphics[width = 0.96\textwidth]{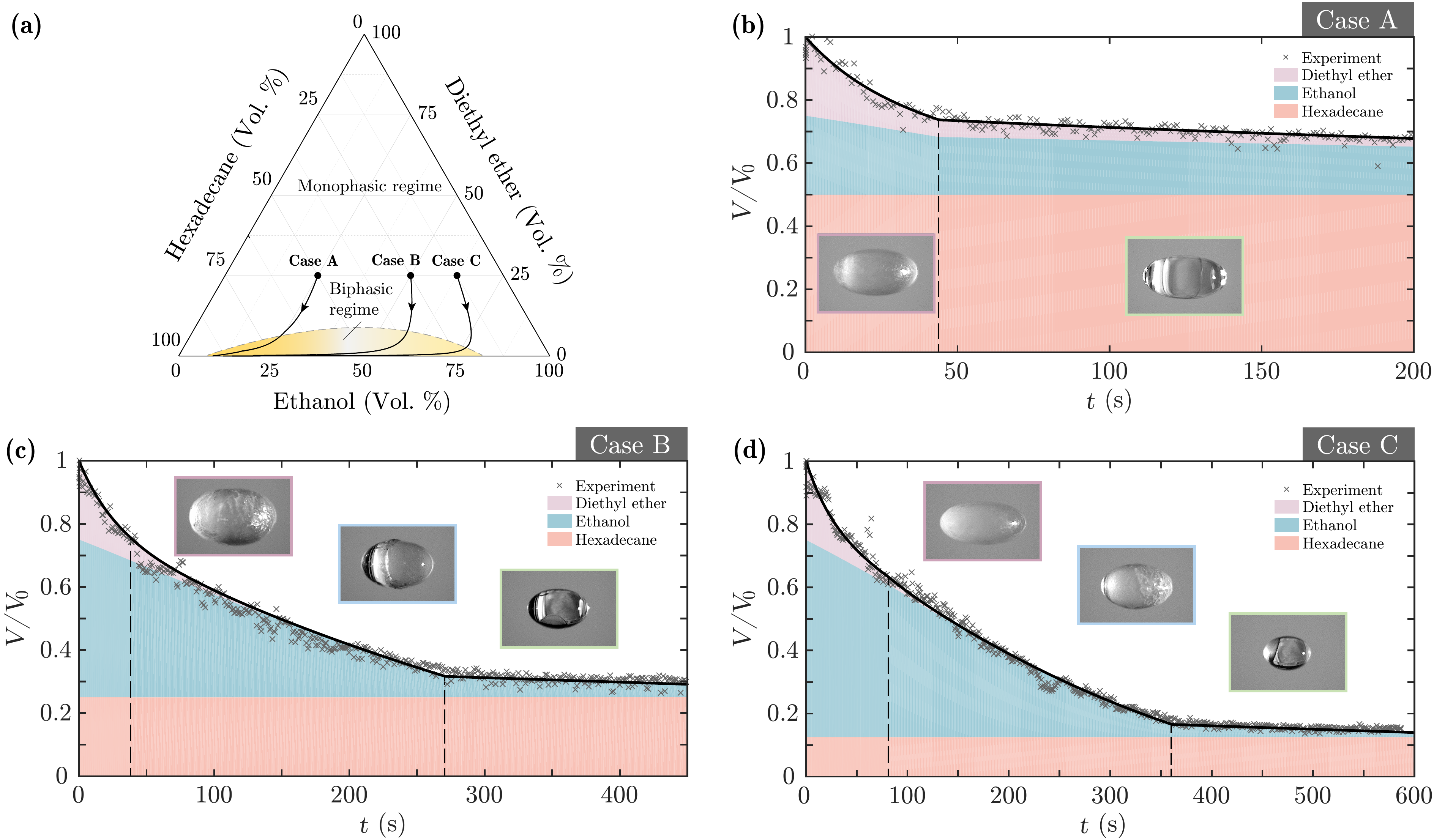}
\caption{Evaporation dynamics for ternary droplets with different initial compositions. (a) Ternary phase diagram for the system of ethanol/hexadecane/diethyl ether in volume fractions at 25$^\circ$. The dashed line represents the miscibility limit between
the single phase domain and the multiphase domain (yellow shaded area, including Ouzo and biphasic regimes). The solid lines represent the predicted compositional paths during evaporation. (b$\sim$d) The evolution of the volume fractions of the three components for the three cases is shown in (a). The volume of each component is normalized by the initial droplet volume. The black crosses represent the experiment data. The solid line represents the model predictions. The vertical dashed lines mark the transition between different stages of evaporation. The insert shows snapshots of representative droplets at different stages of evaporation.}
\label{FIG5}
\end{figure*}

It is apparent from Fig.~\ref{FIG4}(b) that, initially, the volume loss stems predominantly from the evaporation of the most volatile component, namely diethyl ether. The evaporation rate in the initial few seconds is compared to the pure diethyl ether droplet which is suspended in the same acoustic field. As a control experiment, the pure diethyl ether droplet evaporates at an expected rate (pink diamond and pink curve), i.e. a linear relation $L^2 \sim t$. However, the evaporation rate of the ternary droplet deviates from that of pure diethyl ether just after a moment and decreases progressively. This slowdown can be attributed to the concentration reductions of volatile components induced by selective evaporation. Following the same approach, we calculate the mass loss of each individual component content and hence obtain the volume fraction of each component as a function of time, as depicted in Fig.~\ref{FIG4}(c). It is observed that there is a significant mass loss of diethyl ether in the ouzo state, as a result of the selective evaporation, leading to a rapid decrease in volume fraction. As the droplet evaporation proceeds to the middle of the Janus stage, it is almost exhausted from the ternary droplet. Nevertheless, there is not a sharp transition in volume loss between the Ouzo and Janus states. This can be related to the fact that the evaporation of residual diethyl ether still predominates in the early stage 2. Then, ethanol dominates and contributes to evaporation. The evaporation rate is further reduced. The total evaporation flux is the summation of each compartment, one the ethanol-rich phase and one the hexadecane-rich phase. The surface area of each compartment is linearly correlated to the occupied volume ratio in the entire droplet (for detailed derivation please refer to the Supplementary Materials). Due to the minor solubility of ethanol in hexadecane, the volatile components are concentrated in the ethanol-rich part, which results in a significant evaporation heterogeneity within the Janus droplet. In the middle and late stages of the Janus state, the periodic behavior of hexadecane solidification-melting is completely dominated by ethanol evaporation.

Towards the end of the droplet's lifetime, linear regression in droplet volume is observed as the encapsulation of the ethanol-rich phase. A rather sharp transition in the evaporation rate in the second and third stages can be observed (see Fig.~\ref{FIG4}(b)). In this state, the volatile ethanol diffuses from the liquid-liquid interface of encapsulated inner phase to reach the gas-liquid interface. Law and Law~\cite{law1982d2} formulated a multi-component analog of the classical $d^2$-law of droplet evaporation based on the concept of extremely slow mass diffusion in the liquid phase, and the droplet concentration distribution remains almost constant during much of the droplet lifetime. In the modeling work, with regard to the poor solubility of ethanol in the outer phase, here one hypothesis was considered that the concentration gradient of ethanol in the out phase is negligible, thus ensuring a constant and saturated ethanol concentration level in the outer hexadecane phase. The good agreement between the model and experiment validates this hypothesis.


In the current set of experiments, acoustic streaming, temperature gradient caused by evaporation, and rotation of the drop itself would work together to cause complex flows within the droplet. Considering that the evaporation dynamics of the ternary droplet is dependent on the phase transition and diffusion behaviors inside, more experiments on the evaporation of droplet with different initial compositions were conducted. 

As shown in Fig.~\ref{FIG5}(a), three cases with the same diethyl ether content (25$\%$) and different hexadecane/ethanol ratios, are considered. The initial composition of the ternary mixtures at the beginning of the evaporation process is homogeneous. All the droplets released had an initial volume of about 2.4 $\mu L$. Here Case B corresponds to the composition of the ternary droplet we discussed in the previous section. The three-component hexadecane/diethyl ether/ethanol mixtures can outline a rich ternary phase diagram where distinct domains can be distinguished. The miscibility-limit line (the gray dashed line) is the best fit of the experimental points obtained by sequentially adding ethanol to a clear binary solution of hexadecane/diethyl ether and then observing until the metastable Ouzo phenomenon appears. The Ouzo zone is located near the phase boundary and occupies only a small composition range (the characteristic border of this region was not determined in this study). In this diagram, the monophasic domain, extending from the hexadecane-rich part to the diethyl ether-rich part occupies a large portion of this diagram. It corresponds to the mixtures which are macroscopically homogeneous and thermodynamically stable. For these three ternary droplets initially located in the monophasic region, the evolution of compositions during evaporation can be roughly determined with our model (the solid lines). With an increasing amount of ethanol, the evaporation behavior evolves from a path along a constant proportion of ethanol to a path along a constant proportion of hexadecane. 

Furthermore, the corresponding evaporation dynamics for the three cases are obtained and depicted in Fig.~\ref{FIG5}(b$\sim$d). For comparison, the volume of each component is normalized using the initial droplet volume. It is evident that diverse evaporating behaviors are revealed. For case A, the droplet's evaporation rate suddenly slowed down after the first stage (see Fig.~\ref{FIG5}(b)). The visualization experiment shows that, after the solid hexadecane shell melts, the liquid film does not break but completely envelops the inner phase, hence entering the third stage (encapsulation state)~\cite{manev2005critical}. The absence of the Janus state is intuitively attributed to the low ethanol content. The high volume fraction of hexadecane promotes the coalescence of micro-droplets in the Ouzo regime to form a macroscopic sub-phase covering on the droplet surface. By contrast, as the ethanol volume fraction increases to 62.5$\%$, the diethyl ether is almost exhausted in the first stage (Fig.~\ref{FIG5}(d)). Moreover, when the encapsulation configuration is formed, only trace amounts of the volatile component, ethanol, remain in the droplet. More than 80$\%$ of the liquid droplet have been vaporized. Consequently, although the droplet of case C takes longer to form the encapsulation configuration, it is the earliest to end the entire evaporation course (see Supplementary Materials). Therefore, the composition of the ternary droplet plays a vital role in the phase transition and evaporation dynamics, e.g., eliminating the periodic behavior of freezing \& melting, accelerating the evaporation completion, etc.

\section{Conclusions}
Our results demonstrate the multifaceted evaporating dynamics of ternary droplets that are fundamentally different from the typically studied scenario of evaporating single-component or binary droplets\cite{VANGAALEN2020888,chen2022evaporation,tan2016evaporation}. The effects introduced by the selective evaporation of volatile components are highlighted twofold: (1) the composition of ternary droplets crosses the miscibility limit, resulting in spontaneous phase separation in the mixture; (2) the distribution of the precipitated phase evolves with the droplet concentration, giving rise to diversities in droplet configuration and evaporation modes. As the evaporation of volatile components advances, three distinct stages, namely "Ouzo state", "Janus state" and "Encapsulation state", can be identified successively in the evaporation process. With the presence of a high freezing point component, the contact-less freezing induced by evaporation cooling in a ternary droplet is first reported under atmospheric conditions to the best of our knowledge. 
Theoretically, a multistage model is developed by expanding pioneer’s works~\cite{brenn2007evaporation,yarin2002evaporation,mason1967flow}, and further considering the multi-stage evaporation dynamics observed in this study. The results are in quantitative agreement with the experimental observations. We have demonstrated that the frozen solid layer covering the droplet surface can be regarded as a layer of porous media, which changes the acoustic boundary condition and introduces new physical mechanisms. Further instability analysis study is underway to determine the critical concentration that the phase transitions can happen. And more details are needed to consider in the model to describe the transient evaporation dynamics. Our findings can be generically extended to similar ternary solution systems where differences in physicochemical properties of fluid components would trigger unique phase behavior and dynamics. The experiment system itself exhibits the potential to be a good model for studies concerning the evaporation-induced solidification phenomenon in multicomponent droplets. Tuning the droplet's initial composition and chemical properties of individual components may be potential in a variety of applications, such as printing, drug delivery/production, and chemical engineering industries. The introduction of spontaneous evaporative freezing provides even more opportunities for promoting the liquid phase separation and for the non-intrusive manipulation of multiphase configurations in an evaporating liquid system.


\section*{Declaration of Competing Interest}

The authors declare that they have no known competing financial interests or personal relationships that could have appeared to influence the work reported in this paper.

\section*{Acknowledgments}
This work was financially supported by the National Natural Science Foundation of China under Grant Nos. 11988102, 12202244, and 51976105, and Tencent Foundation through the XPLORER PRIZE. We are grateful to Wen Li, Jingtao Liu and Yong Gao for the helpful conversations.

\providecommand{\noopsort}[1]{}\providecommand{\singleletter}[1]{#1}%

\end{document}